\documentclass[prl,twocolumn,amsmath,amssymb,showpacs]{revtex4}
\usepackage{graphicx}
\usepackage{dcolumn}
\usepackage{bm}
\begin{document}
\newcommand{\widthfigA}{0.35\textwidth} 
%
%
%
\title{Time-oscillating Lyapunov modes and 
auto-correlation functions 
for quasi-one-\\dimensional systems} 
\author{Tooru Taniguchi and Gary P. Morriss}
\affiliation{School of Physics, University of New South Wales, Sydney, 
New South Wales 2052, Australia}
\date{\today}
\begin{abstract}
   The time-dependent structure of the Lyapunov vectors 
corresponding to the steps of Lyapunov spectra 
and their basis set representation  
are discussed for a quasi-one-dimensional many-hard-disk systems. 
   Time-oscillating behavior is observed in 
two types of Lyapunov modes, 
one associated with the time translational invariance and 
another with the spatial translational invariance, 
and their phase relation is specified. 
   It is shown that the longest period of 
the Lyapunov modes is twice as long as the period of 
the longitudinal momentum auto-correlation function. 
   A simple explanation for this relation is proposed. 
   This result gives the first quantitative connection between
the Lyapunov modes and an experimentally accessible quantity.  
\end{abstract}
\maketitle  
%
%
%
   In recent years, statistical mechanics based on chaotic 
dynamics has drawn much attention as a scheme to bridge between 
a deterministic microscopic dynamics and a probabilistic 
macroscopic theory. 
   In a chaotic system, the difference of two nearby 
phase space trajectories, the so called Lyapunov vector, 
diverges rapidly in time, so that a part of the dynamics is 
unpredictable and a statistical treatment is required. 
   To justify this scheme, much effort has been devoted 
to discovering the link between macroscopic quantities, 
such as transport coefficients, 
and chaotic quantities, such as
the Lyapunov vector and the Lyapunov exponent (the 
exponential rate of divergence or contraction of the 
magnitude of the Lyapunov vector) \cite{Gas98,Dor99}. 
   Some works have concentrated upon finding chaotic properties 
of many-particle systems, and 
some progress has been made in the search for global properties,
like the conjugate pairing rule for the Lyapunov spectrum 
(the ordered set of Lyapunov exponents) 
\cite{Dre88}.

   The stepwise structure of the Lyapunov spectrum is 
a typical chaotic feature of many-particle systems \cite{Del96}. 
   These steps in the spectrum are accompanied by global wavelike 
structures in the corresponding Lyapunov vectors, 
or Lyapunov modes \cite{Pos00,Tan03a,For04,EF04}.  
   The significance of this phenomenon is that this structure 
appears in the vectors associated with the  
Lyapunov exponents that are closest to zero, 
therefore it is connected with the slow 
macroscopic behavior of the system. 
   Originally, it was observed in systems with 
hard-core particle interactions, but very recently 
numerical evidence for Lyapunov modes in 
a system with soft-core interactions was reported \cite{Yan04}. 
   Some analytical approaches, such as random matrix 
theory \cite{Eck00,Tan02c}, kinetic theory 
\cite{Mcn01b} and periodic orbit theory \cite{Tan02b}, 
have been used in an attempt to understand this phenomenon. 

   A clue to understanding the stepwise structure of the Lyapunov 
spectrum and the Lyapunov modes is in the behavior of the 
Lyapunov vectors associated with the zero-Lyapunov exponents.  
   In this scenario we claim that the origin of the Lyapunov modes is the 
translation invariances and conservation laws  
which dominate the global behavior the system. 
   For a two-dimensional system of 
$N$ particles in periodic boundary conditions   
the Lyapunov vectors can be written as 
$(\delta x, \delta y, \delta p_{x}, \delta p_{y})$
where $\delta x$,  $\delta y$, $\delta p_{x}$ and  $\delta p_{y}$
are the $N$-vectors containing the $x$ and $y$ spatial components,
and the $x$ and $y$ momentum components
of the Lyapunov vector. 
For this system there are six degenerate zero-Lyapunov exponents
and the Lyapunov modes are 
linear combinations of one of the basis sets 
$\{ (\Delta, 0, 0, 0)$, 
$(0, \Delta, 0, 0)$, 
$( p_{x},  p_{y}, 0, 0) \}$, 
or
$\{(0, 0, \Delta ,0)$,  
$(0, 0,0, \Delta )$,  
$(0, 0, p_{x},  p_{y}) \}$,
where $0$ is the $N$-dimensional null vector, 
$\Delta$ is an $N$-dimensional vector with all components 
equal to $1/\sqrt{N}$,
and $p_{x}$ and $p_{y}$ are the $N$-dimensional vectors containing
 the $x$ and $y$-components of the momentum of each particle \cite{Gas98}. 
  Notice that the second basis set is the conjugate of the first,
where $(-\delta p, \delta q) $ is the conjugate of $(\delta q, \delta p ) $. 
  The first set of basis vectors are associated with 
spatial and time translational invariance, the second basis set  
with total momentum and energy conservation. 
   Therefore, in the zero-Lyapunov modes we observe that the sets 
$\{\delta x_{j}^{(n)}\}$, $\{\delta y_{j}^{(n)}\}$, 
$\{\delta p_{xj}^{(n)}\}$, $\{\delta p_{yj}^{(n)}\}$, 
$\{\delta x_{j}^{(n)}/p_{xj}, \delta y_{j}^{(n)}/p_{yj}\}$ or
$\{\delta p_{xj}^{(n)}/p_{xj}, \delta p_{yj}^{(n)}/p_{yj}\}$ 
can have constant components independent of the particle number index $j$ \cite{Tan03a}.    
   If the zero-Lyapunov exponents and their  Lyapunov vectors  
can be regarded as the zero-th step 
then we would expect to see $x$ dependent Lyapunov modes for $k$-vector $k_{n}=2 n \pi /L_{x}$ 
constructed from the basis vectors
\begin{eqnarray}
(0, \sin(k_{n} x_{j}), 0, 0), 
(0, \cos(k_{n} x_{j}), 0, 0),   \label{Ts&c} \\
(\sin(k_{n} x_{j}), 0, 0, 0), 
(\cos(k_{n} x_{j}), 0, 0, 0),   \label{Ls&c} \\
(p_{xj} \sin(k_{n} x_{j}),  p_{yj} \sin(k_{n} x_{j}), 0, 0),   \nonumber \\
(p_{xj} \cos(k_{n} x_{j}),  p_{yj} \cos(k_{n} x_{j}), 0, 0),   \label{Lps&c} 
\end{eqnarray}
in the quasi-one-dimensional system with periodic boundary conditions \cite{Tan03a}. 
   However, only the first two of these (\ref{Ts&c})
are observed (as the two-point steps or transverse modes) \cite{Tan03a, EF04}.
   The second two modes are longitudinal and we notice that the last 
two basis vectors (\ref{Lps&c}) have time dependent normalisation coefficients.
   We can remove this time dependence by combining the time-translational invariance 
mode and the longitudinal modes in a particular way.
   It is these combined modes that form an approximate basis set for
the numerically observed Lyapunov modes 
\begin{eqnarray}
\alpha \sin(\omega_{n} t)(p_{xj} \sin(k_{n} x_{j}),  p_{yj} \sin(k_{n} x_{j}), 0, 0 )    \nonumber \\
       + \beta \cos(\omega_{n} t) ( \cos(k_{n} x_{j}) , 0, 0, 0 ),   \label{Lss} \\
\alpha \sin(\omega_{n} t)(p_{xj} \cos(k_{n} x_{j}),  p_{yj} \cos(k_{n} x_{j}), 0, 0) \nonumber \\  
       + \beta \cos(\omega_{n} t) ( \sin(k_{n} x_{j}) , 0, 0, 0 ),   \label{Lsc}  \\
\alpha  \cos(\omega_{n} t) (p_{xj} \sin(k_{n} x_{j}), p_{yj} \sin(k_{n} x_{j}),  0, 0 ) \nonumber\\
       + \beta \sin(\omega_{n} t) ( \cos(k_{n} x_{j}), 0, 0, 0 ),   \label{Lcs}  \\
\alpha  \cos(\omega_{n} t) (p_{xj} \cos(k_{n} x_{j}), p_{yj} \cos(k_{n} x_{j}),0, 0 )  \nonumber\\
        + \beta \sin(\omega_{n} t) ( \sin(k_{n} x_{j}) , 0, 0, 0 ),   \label{Lcc} 
\end{eqnarray}
(where $\omega_{n}$ is a frequency) which are normalisable at all times.

   Although this scenario for the Lyapunov steps and modes 
seems to be plausible, convincing evidence is still absent. 
   The wavelike structure in $\delta y_{j}^{(n)}$ as a 
function of $x_{j}$, the so called the transverse spatial 
translational invariance Lyapunov mode, is well known \cite{Pos00}. 
   This mode is stationary in time, and categorizes 
one of the two types of Lyapunov steps. 
   Ref. \cite{Tan03a} showed a mode structure in 
$\delta y_{j}^{(n)}/p_{yj}$ as a function of $x$, namely 
the transverse time translational invariance Lyapunov mode. 
   This mode appears as a time-oscillating spatial wavelike 
structure, and categorizes another type of Lyapunov step. 
   On the other hand, modes of the 
longitudinal components of the Lyapunov vectors are less 
convincing. 
   Ref. \cite{For04} claims a moving mode structure 
in $\delta x_{j}^{(n)}$ as a 
function of $x$, in the same Lyapunov steps as the ones having 
the mode in $\delta y_{j}^{(n)}/p_{yj}$, but the relation between 
these two modes are not known. 
   Another important unsolved problem is the time scale 
of the oscillation of the Lyapunov mode. 
   The period is rather long, 
often thousands of times as long as the mean free time, 
and it should correspond to a macroscopic 
or collective property of the system. 
   However, no direct evidence of this has been reported. 

   In this letter we show explicit numerical evidence 
for the time-oscillating wavelike structure of the 
multiple longitudinal Lyapunov modes, 
in particular $\delta x_{j}^{(n)}/p_{xj}$ and $\delta x_{j}^{(n)}$ 
as functions of $x_{j}$ [basis vectors (\ref{Lss}), (\ref{Lsc}), (\ref{Lcs}), (\ref{Lcc})],
and clarify the phase relation between them and the mode $\delta y_{j}^{(n)}/p_{yj}$. 
   Further, we present numerical evidence to 
connect the time-oscillation of the Lyapunov modes 
to the time-oscillation of the momentum auto-correlation function. 
   Oscillatory behavior has been observed for the momentum 
auto-correlation function in many macroscopic systems 
\cite{Rah67,Han90}, 
and has its origin in a collective motion of the system 
\cite{Zwa67}. 
   Our central result is that the longest period of oscillation 
of the Lyapunov modes is twice as long as the period of the 
longitudinal momentum auto-correlation function. 
   Although the period changes with boundary conditions and with 
the number of particles $N$, this inter-relation between periods 
is always satisfied. 
   A possible explanation for this relation is proposed. 
   The auto-correlation function can be observed directly by the 
neutron and light scattering experiments \cite{Han90,CL75} 
and is connected to transport coefficients using linear 
response theory \cite{Kub78}, 
so this result gives quantitative evidence of 
a connection between the Lyapunov 
modes and an experimentally accessible quantity. 

   One of the difficulties of observing the Lyapunov steps and modes 
is that numerical calculations of Lyapunov spectra 
and vectors are very time-consuming, 
and analytical methods for them are not well established, 
although there have been some attempts \cite{Tan02c,Tan02b,New86}. 
   On the other hand, if the above scenario for the 
Lyapunov steps and modes, based on universal properties 
like translational invariance and the conservation laws,  
can be justified, then a simple model should be sufficient to 
establish its origin. 
   This is the motivation behind the quasi-one-dimensional 
many-hard-disk system proposed by
\cite{Tan03a,Tan03b}. 
   It is a very narrow rectangular system where the 
particle ordering in the longer $x$-direction is invariant. 
   Particle interactions are restricted to 
nearest neighbors only, so the numerical calculation 
is much faster than for the full two-dimensional system. 
   Moreover it shows a much longer region of Lyapunov steps 
and clearer Lyapunov modes, compared with the full 
two-dimensional system. 
   Another useful technique to get clear Lyapunov steps and modes, 
is to use hard-wall boundary conditions \cite{Tan03a}. 
   Although hard-wall boundary conditions break  
translational invariance and momentum conservation, 
there is a simple relationship 
between the Lyapunov steps and modes for systems 
with different boundary conditions \cite{Tan03a}. 
   In this letter we use the 
quasi-one-dimensional many-hard-disk system with hard-wall 
boundary conditions in the longer $x$ direction and 
periodic boundary conditions in the shorter $y$ direction 
[the boundary condition (H,P)]. 
   This system exhibits both types of Lyapunov steps, 
and is only marginally slower numerically 
because only the two particles at the ends of the system 
can collide with the walls. 
   We use units where  
the mass and disk radius $R$ of each particle is $1$, the 
total energy is $N$ and 
the dimensions of the system are chosen as 
$L_{x} = 1.5 N L_{y} + 2R$ and $L_{y} = 2R(1+10^{-6})$, respectively. 
   
\begin{figure}[t]
\includegraphics[width=\widthfigA]{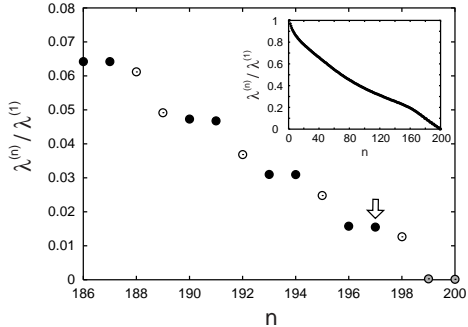}
\caption{
      Stepwise structure of the Lyapunov spectrum 
   normalized by the largest Lyapunov exponent $\lambda^{(1)}$
   for the quasi-one-dimensional many-hard-disk system 
   with (H,P) boundary condition. 
      The number of particles is $N=100$.
      The white-filled circles correspond 
   to the stationary modes while the black filled circles 
   are the time-oscillating modes. 
      Inset: The positive branch of the normalized Lyapunov spectrum. 
   }
\label{fig1lyapuStep}\end{figure}  

   Figure \ref{fig1lyapuStep} shows the stepwise structure 
of the normalized Lyapunov spectrum for the quasi-one-dimensional 
system consisting of $100$ hard-disks with (H,P) boundary condition. 
   In the inset to this figure we show the full normalized spectrum 
for the positive branch (the negative branch is obtained by the conjugate 
pairing rule). 
   In this system the spatial translational invariance 
and the total momentum conservation in the longitudinal 
direction are violated, and the number of the zero-Lyapunov 
exponents is $4$.  
   We recognize a clear stepwise structure in which 
two-point steps and one-point steps appear alternately.

\begin{figure}[t]
\includegraphics[width=\widthfigA]{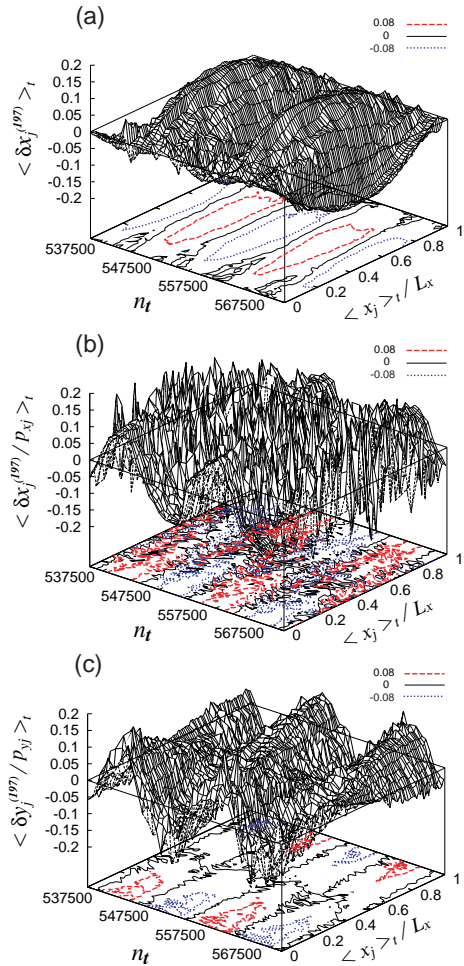}
\caption{
      Local time averages of: 
 (a)  $\langle \delta x_{j}^{(197)} \rangle_{t}$; 
 (b) $\langle \delta x_{j}^{(197)}/p_{xj} \rangle_{t}$;  and
 (c)  $\langle \delta y_{j}^{(197)}/p_{yj} \rangle_{t}$;  
   as functions of the collision number $n_{t}$ and 
   the normalized position $\langle x_{j} \rangle_{t}/L_{x}$ of 
   the $x$-component of the $j$-th particle. 
      The corresponding Lyapunov exponent is indicated by the 
   arrow in Fig. \ref{fig1lyapuStep}. 
      On the bottom of each graph is the projected contour plot of the 
   three dimensional graph at the values $-0.08$ (dotted lines), 
   $0$ (solid lines), and $+0.08$ (broken lines). 
   }
\label{fig2lyaMode}\end{figure}  

   One-point steps in the Lyapunov spectrum 
in Fig. \ref{fig1lyapuStep} correspond to the  
stationary mode $\delta y_{j}^{(n)}$ 
as a function of $x_{j}$ [basis vectors (\ref{Ts&c})]. 
   This structure is well known as the transverse 
spatial translational invariant Lyapunov mode,  
so we omit further discussion here. 
   Figure \ref{fig2lyaMode} shows that 
the Lyapunov modes corresponding to the two-point 
Lyapunov steps have time-oscillating wavelike mode 
structures in $\delta x_{j}^{(n)}/p_{xj}$, 
$\delta y_{j}^{(n)}/p_{yj}$ and $\delta x_{j}^{(n)}$ 
as functions of $x_{j}$ and the collision number $n_{t}$ 
with almost the same period 
$T_{lya} \approx 16300$. 
   More details of this numerical calculation are given 
elsewhere \cite{Tan03a,Tan04b}. 
   By investigating time-dependent Lyapunov modes 
for the other Lyapunov exponents \cite{Tan04b}, it is shown  
that the spatial part of Lyapunov vector 
components $\delta x$ and $\delta y$ corresponding to the 
Lyapunov exponents in the $n$-th two-point step 
are approximately expressed using the basis vectors (\ref{Lsc}) and (\ref{Lcc}).
   Notice however, that for (H,P) boundary conditions $k_{n} 
=n\pi /L_{x}$. 
   Note that the time-translational invariance Lyapunov modes, 
that is the terms containing  the momentum in basis vectors
(\ref{Lss}), (\ref{Lsc}), (\ref{Lcs}) and (\ref{Lcc}) 
have the same structure in the same Lyapunov exponent, 
and are orthogonal to the longitudinal spatial 
translational invariant Lyapunov components of the same
basis vectors. 
   Large fluctuations in the longitudinal time translational 
invariance Lyapunov mode, like in Fig. \ref{fig2lyaMode}(b),  
come from the terms for the 
longitudinal spatial translational invariance Lyapunov modes.
   The spatial nodes of these Lyapunov modes are {\it pinned} 
at the hard-walls \cite{Tan03a,Tan04b}.

\begin{figure}[t]
\includegraphics[width=\widthfigA]{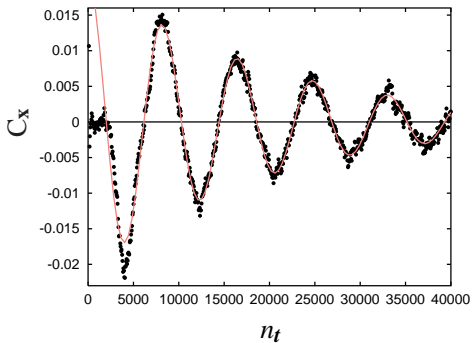}
\caption{
      The time-oscillating part of the longitudinal 
   momentum auto-correlation function 
   $C_{x}$,  normalized by its initial value, 
   as a function of the collision number $n_{t}$. 
      The system is the same as that used in Figs. 
   \ref{fig1lyapuStep} and \ref{fig2lyaMode}. 
      Here, the line is a fit of the data to a sinusoidal function 
   multiplied by an exponential function. 
   }
\label{fig3AcfOscill}\end{figure}  

   We now discuss the connection between the  
time-oscillations of Lyapunov modes and 
the momentum auto-correlation function. 
   Figure \ref{fig3AcfOscill} shows the longitudinal 
momentum auto-correlation normalized by its initial value 
as a function of the collision number $n_{t}$ in 
the same system as that used in Figs. 
\ref{fig1lyapuStep} and \ref{fig2lyaMode}.
   Here, an arithmetic average of the auto-correlation functions 
for $11$ disks in the middle of the system (far from the hard walls) 
is taken.
   This auto-correlation function initially shows a simple exponential 
damping, as discussed elsewhere \cite{Tan04b}, 
so that part is omitted in Fig. \ref{fig3AcfOscill}.  
   A clear time-oscillating behavior is observed in 
Fig. \ref{fig3AcfOscill} that is nicely fitted to a sinusoidal function 
times an exponential decay. 
   This fitting gives the oscillating period $T_{acf} 
\approx 8290$, which suggests the approximate relation  
\begin{eqnarray}
   T_{lya} \approx 2 T_{acf}  
\label{RelationT}\end{eqnarray}
\noindent between the 
oscillating periods of the Lyapunov mode and the momentum 
auto-correlation function. 
   Numerically this relation is 
satisfied irrespective of the number of particles and  
boundary conditions \cite{Tan04b}. 

   The relation (\ref{RelationT}) can be obtained from  
the following physical argument. 
   Consider the Fourier space auto-correlation function 
$<\tilde{p}_{x}(t)^{*}\;  \tilde{p}_{x}(0) >$ for 
the longitudinal component $\tilde{p}_{x}(t)$ of the momentum 
that oscillates with frequency $\omega_{acf}$, namely 
$
   <\tilde{p}_{x}(t)^{*} \; \tilde{p}_{x}(0)>
   \sim \phi(t) \; \exp\{i\omega_{acf} t\},
$
where 
$\phi(t)$ is the damping factor 
and the suffix ${}^{*}$ indicates the complex conjugate. 
   On the other hand we have a term for the 
longitudinal time translational invariance Lyapunov mode, 
like the first component of the basis vectors 
(\ref{Lsc}) and (\ref{Lcc}), as 
$
   \delta\tilde{q}_{x} \sim 
   \psi_{1}(t) \tilde{p}_{x}(t) \exp\{i\omega_{lya}t\}
$   
with a frequency $\omega_{lya}$, 
where $\psi_{1}(t)$ is an amplitude factor.
   We assume that if the quantity 
$\delta\tilde{q}_{x}$ oscillates persistently in time, 
then its auto-correlation function 
$<\delta\tilde{q}_{x}(t)^{*} \; \delta\tilde{q}_{x}(0)>$ 
should also oscillate in time 
with the same frequency $\omega_{lya}$, namely $
 <\delta\tilde{q}_{x}(t)^{*} \;\delta\tilde{q}_{x}(0)> 
  \sim \psi_{2}(t) \exp\{i\omega_{lya}t\} $ with a damping factor $\psi_{2}(t)$. 
   (Actually we can show that the auto-correlation function 
for the spatial longitudinal component of the Lyapunov vector 
oscillates with the same frequency as the 
Lyapunov vector component itself in the 
quasi-one-dimensional system \cite{Tan04b}.) 
   It follows from these equations and the assumption that 
$
   \psi_{2}(t) \exp\{i\omega_{lya} t\}
      \sim \psi_{1}(t)^{*}\;\psi_{1}(0)  \; \phi(t) \;  
      \exp\{i (\omega_{acf} - \omega_{lya})t \} 
$, 
that $\omega_{lya} \sim \omega_{acf} /2$. 
   The frequency is inversely proportional to 
the oscillation period, 
so we obtain relation (\ref{RelationT}). 


   In conclusion, we have discussed time-oscillating Lyapunov modes 
and the momentum auto-correlation function in quasi-one-dimensional 
many-hard-disk systems. 
   We clarified the phase relation between components of multiple 
time-dependent Lyapunov modes, one coming from the longitudinal spatial 
translational invariance and another from the time translational 
invariance.
   We showed that the longest time-oscillating 
period of the Lyapunov modes is almost twice as long as 
the time-oscillating period of the longitudinal momentum 
auto-correlation function. 
   The momentum auto-correlation can be measured through 
experiments like the neutron and light scattering experiments, so 
our result give a direct connection between the Lyapunov mode 
with an experimentally accessible quantity.
 
   We are grateful for the financial support for this work 
from the Australian Research Council. 
   One of the authors (T. T.) also appreciates the financial 
support by the Japan Society for the Promotion Science. 

%
%

%
\end{document}